\documentstyle[seceq,preprint]{ptptex}

\newcommand{\definition}{\stackrel{\rm d}{\equiv}}
\newcommand{\second}{I\hspace{-1mm}I} 

\preprintnumber{TOYAMA-91
}
\title{
Gauge Theory of Massive Tensor Field ~\second \\
{\it --- Covariant Expressions ---}
}

\author{
Shinji {\sc Hamamoto}
\footnote{E-mail address: hamamoto@sci.toyama-u.ac.jp}
}

\inst{
Department of Physics, Toyama University, Toyama 930
}

\recdate{
\today
}

\abst{
Covariant forms are given to a gauge theory of massive tensor 
field.
This is accomplished by introducing another auxiliary field of 
scalar type to the system composed of a symmetric tensor field and 
an auxiliary field of vector type. 
The situation is compared to the case of the theory in which a 
tensor field describes a scalar ghost as well as an ordinary 
massive tensor. 
In this case only an auxiliary vector field is needed to give 
covariant expressions for the gauge theory. 
}

\begin{document}

\maketitle

\section{Introduction}
In a previous paper \cite{rf:1}(referred to as I) a massive tensor 
field theory with a smooth massless limit was constructed. 
We applied the Batalin-Fradkin (BF) algorithm \cite{rf:2} to the 
pure-tensor (PT) model which describes a massive pure tensor of 
five degrees of freedom. 
By introducing an auxiliary vector field, we converted the original 
second-class constrained system into a first-class one. 
To the gauge-invariant system we obtained, massless-regular 
gauge-fixing was imposed. 
The resulting theory was found to have a smooth massless limit. 
Based on the Hamiltonian formalism, however, our formulation is 
non-covariant from the beginning. 
The final result has been left lacking in covariance. 

The purpose of the present paper is to give covariant forms to 
the result. This is accomplished by introducing another auxiliary 
field of scalar type in addition to the auxiliary field of vector 
type. 
The situation is to be compared to the case of the 
additional-scalar-ghost (ASG) model where a symmetric tensor field 
describes an additional scalar ghost as well as the ordinary 
massive tensor. 
For the ASG model, an auxiliary vector field has also to be 
introduced to convert the original second-class constrained system 
into a first-class one. 
In this case, however, this is enough. 
It is seen that we can obtain covariant expressions without 
introducing any other auxiliary field.

In \S 2, canonical formalism of massive tensor field is presented. 
It is shown that the structures of constraints are different 
according to the value of a parameter $a$ in mass terms. 
When $a = 1$, which gives the PT model, we have five kinds of 
constraints, four second-class and one first-class. 
On the other hand, in the case of $a \not= 1$, which corresponds to 
the ASG model, we have only four kinds of second-class constraints. 
In \S 3, the BF algorithm is applied to these systems. 
For both cases of $a = 1$ and $a \not= 1$, we can convert all the 
second-class constraints to first-class ones by introducing an 
auxiliary field of vector type.
In \S 4, we investigate massless-regular gauge-fixings that allow 
to take smooth massless limits. 
Covariant expressions for the final results are given in \S 5. 
In the case of $a \not= 1$, we can easily find covariant path 
integral expressions. 
When $a = 1$, however, we have to introduce another auxiliary field 
of scalar type in order to write down the result in covariant 
forms. 
Section 6 gives summary.

\section{Canonical formalism}

A massive tensor field is described by the Lagrangian
\footnote
{
In the present paper Greek indices run $0-3$, 
while Latin indices $1-3$.
The metric is 
$\eta^{\mu\nu} \definition ( -1, +1, +1, +1 )$.
}
\begin{equation}
L[h] = L[h, m=0] 
- \frac{m^{2}}{2}\left( h_{\mu\nu}h^{\mu\nu} - ah^{2}\right) ,
\end{equation}
where $L[h, m=0]$ represents the Lagrangian for a massless tensor 
field
\begin{equation}
L[h, m=0] \definition 
- \frac{1}{2}\left( 
\partial_{\lambda}h_{\mu\nu}\partial^{\lambda}h^{\mu\nu} 
- \partial_{\lambda}h\partial^{\lambda}h\right) 
+ \partial_{\lambda}h_{\mu\nu}\partial^{\nu}h^{\mu\lambda} 
- \partial_{\mu}h^{\mu\nu}\partial_{\nu}h ,
\end{equation}
$h$ stands for the trace of $h_{\mu\nu}$ 
$\left( h \definition h^{\mu}_{\makebox[1mm]{}\mu}\right)$, 
and $a$ is a real parameter.
Field equations are
\begin{eqnarray}
\left(\Box - m^{2}\right) h_{\mu\nu} 
- (2a-1)\left(\partial_{\mu}\partial_{\nu} 
+ \frac{m^{2}}{2}\eta_{\mu\nu}\right) h & = & 0 , \\
\partial^{\nu}h_{\mu\nu} - a\partial_{\mu}h & = & 0 , \\
2(a-1)\Box h + (4a-1)m^{2}h & = & 0 .
\end{eqnarray}
For $a=1$ (PT model), the field equations reduce to 
\begin{eqnarray}
\left(\Box - m^{2}\right) h_{\mu\nu} & = & 0 , \\
\partial^{\nu}h_{\mu\nu} & = & 0 , \\
h & = & 0 ,
\end{eqnarray}
which show that this case purely describes a massive tensor field 
with five degrees of freedom.
For other arbitrary value of $a\not= 1$ (ASG model), the Lagrangian 
describes an additional scalar ghost as well as the ordinary tensor 
field. 
In particular $a=\frac{1}{2}$ gives simple field equations 
\begin{eqnarray}
\left(\Box - m^{2}\right) h_{\mu\nu} & = & 0 , \\
\partial^{\nu}h_{\mu\nu} - \frac{1}{2}\partial_{\mu}h & = & 0 .
\end{eqnarray}
To investigate the structure of constraints and Hamiltonian, 
we have to consider two cases of $a \not= 1$ and $a=1$ separately.

\subsection{The case of $a \not= 1$}
In this case we have two primary constraints
\begin{eqnarray}
\varphi^{0} & \definition & 
\pi^{0} \approx 0 , 
\label{eq:211} \\
\varphi^{m} & \definition & 
\pi^{m} + \partial_{n}h^{mn} - \partial^{m}h' \approx 0 , 
\label{eq:212}
\end{eqnarray}
and two secondary constraints
\begin{eqnarray}
\varphi_{1}^{0} & \definition & 
\partial_{m}\partial^{m}h' 
- \partial_{m}\partial_{n}h^{mn} 
- m^{2}\left[ (1-a)h_{0} + ah'\right] \approx 0 , \\
\varphi_{1}^{m} & \definition & 
\partial_{n}\pi^{mn} 
+ \frac{1}{2}\partial_{n}
\left(\partial^{m}h^{n} - \partial^{n}h^{m}\right) 
+ m^{2}h^{m} \approx 0 ,
\end{eqnarray}
where $\pi^{0}$, $\pi^{m}$ and $\pi^{mn}$ are momenta conjugate to 
$h_{0} \definition h_{00}$, $h_{m} \definition h_{0m}$ and 
$h_{mn}$ respectively, and $h'$ denotes the three-dimentional 
trace $h' \definition h^{m}_{\makebox[1mm]{}m}$.
The Poisson brackets between these constraints are calculated as
\begin{eqnarray}
\left[ \varphi^{0}(\mib{x}), \varphi_{1}^{0}(\mib{x}')\right] 
& = & (1-a)m^{2}\delta^{3}(\mib{x}-\mib{x}') , 
\label{eq:215} \\
\left[ \varphi^{m}(\mib{x}), \varphi_{1}^{n}(\mib{x}')\right] 
& = & - m^{2}\eta^{mn}\delta^{3}(\mib{x}-\mib{x}') , 
\label{eq:216} \\
\left[ \varphi_{1}^{0}(\mib{x}), \varphi_{1}^{m}(\mib{x}')\right] 
& = & am^{2}\partial^{m}\delta^{3}(\mib{x}-\mib{x}') , 
\label{eq:217} \\
\mbox{The others} & = & 0 . \nonumber
\end{eqnarray}
The Hamiltonian is 
\begin{equation}
H = H_{0} 
+ \left[\frac{a}{2(1-a)}\pi' 
+ \partial^{m}h_{m}\right]\varphi^{0} 
+ \left[\partial^{n}h_{mn} 
+ a\partial_{m}\left( h_{0}-h'\right)\right]\varphi^{m} ,
\end{equation}
where
\begin{equation}
H_{0} \definition 
H_{0}(m=0) 
+ \frac{m^{2}}{2}\left[
(1-a)h_{0}^{\makebox[1mm]{}2} - 2h_{m}h^{m} 
+ h_{mn}h^{mn} + 2ah_{0}h' - ah'^{2}\right] ,
\end{equation}
and
\begin{eqnarray}
H_{0}(m=0) & \definition & 
\frac{1}{2}\pi^{mn}\pi_{mn} - \frac{1}{4}\pi'^{2}
+ \pi^{mn}\partial_{m}h_{n} \nonumber \\ 
& & 
- \frac{3}{4}\left[\partial_{m}h_{n}\partial^{m}h^{n} 
- \left(\partial^{m}h_{m}\right)^{2}\right] 
+ \partial_{m}h_{0}\partial^{m}h' 
- \partial_{m}h_{0}\partial_{n}h^{mn} \nonumber \\
& & 
+ \frac{1}{2}\left( \partial_{l}h_{mn}\partial^{l}h^{mn} 
- \partial_{l}h'\partial^{l}h'\right) 
- \partial_{l}h_{mn}\partial^{n}h^{ml} 
+ \partial_{m}h^{mn}\partial_{n}h' . \nonumber \\
& & \label{eq:220}
\end{eqnarray}
The Poisson brackets between the constraints and the Hamiltonian 
are 
\begin{eqnarray}
\left[ \varphi^{0}, H \right] & = & 
\varphi_{1}^{0} + a\partial_{m}\varphi^{m} , \\
\left[ \varphi^{m}, H \right] & = & 
2\varphi_{1}^{m} 
+ \frac{1-2a}{1-a}\partial^{m}\varphi^{0} , \\
\left[ \varphi_{1}^{0}, H \right] & = & 
- \partial_{m}\varphi_{1}^{m} 
+ \frac{a}{2(1-a)}\left( 2\partial_{m}\partial^{m} 
- 3am^{2}\right)\varphi^{0} , \\
\left[ \varphi_{1}^{m}, H \right] & = & 
\frac{1}{2}\partial_{n}\partial^{n}\varphi^{m} + 
\frac{1}{2}(1-2a)\partial^{m}\partial_{n}\varphi^{n} .
\end{eqnarray}
Equations (\ref{eq:215}), (\ref{eq:216}) and (\ref{eq:217}) 
show that all the four constraints are of the second class.

\subsection{The case of $a=1$}
This case was studied in I. The results are quoted here for the 
sake of comparison with the case of $a \not= 1$.
\footnote
{Some of the equations have minor differences from the 
corresponding ones in I. Since the differences come from total 
divergences in the Lagrangian, however, they have no essential 
effects.} 
Primary constraints are the same as 
(\ref{eq:211}) and (\ref{eq:212}).
For secondary constraints, however, we have three in this case 
\begin{eqnarray}
\varphi_{1}^{0} & \definition & 
\partial_{m}\partial^{m}h' - \partial_{m}\partial_{n}h^{mn} - 
m^{2}h' \approx 0 , \\
\varphi_{1}^{m} & \definition & 
\partial_{n}\pi^{mn} 
+ \frac{1}{2}\partial_{n}\left(\partial^{m}h^{n} 
- \partial^{n}h^{m}\right) + m^{2}h^{m} 
\approx 0 , \\
\varphi_{2}^{0} & \definition & 
\pi' \approx 0 .
\end{eqnarray}
The Poisson brackets are 
\begin{eqnarray}
\left[ \varphi^{m}(\mib{x}), \varphi_{1}^{n}(\mib{x}')\right] 
& = & - m^{2}\eta^{mn}\delta^{3}(\mib{x}-\mib{x}') , \\
\left[ \varphi_{1}^{0}(\mib{x}), \varphi_{1}^{m}(\mib{x}')\right] 
& = & m^{2}\partial^{m}\delta^{3}(\mib{x}-\mib{x}') , \\
\left[ \varphi^{m}(\mib{x}), \varphi_{2}^{0}(\mib{x}')\right] 
& = & - 2\partial^{m}\delta^{3}(\mib{x}-\mib{x}') , \\
\left[ \varphi_{1}^{0}(\mib{x}), \varphi_{2}^{0}(\mib{x}')\right] 
& = & 
\left( 2\partial_{m}\partial^{m} - 3m^{2} \right)
\delta^{3}(\mib{x}-\mib{x}') , \\
\mbox{The others} & = & 0 . \nonumber
\end{eqnarray}
The Hamiltonian is 
\begin{equation}
H = H_{0} + \lambda_{0}\varphi^{0} 
+ \partial^{n}h_{mn}\varphi^{m} + 
( h_{0} - h')\varphi_{1}^{0} ,
\end{equation}
where 
\begin{equation}
H_{0} \definition 
H_{0}(m=0) + \frac{m^{2}}{2}\left( 
- 2h_{m}h^{m} + h_{mn}h^{mn} + 2h_{0}h'- h'^{2}\right) 
\end{equation}
with $H_{0}(m=0)$ defined by (\ref{eq:220}), 
and $\lambda_{0}$ is an arbitrary coefficient.
The Poisson brackets between the constraints and the Hamiltonian 
are
\begin{eqnarray}
\left[ \varphi^{0}, H \right] & = & 0 , \\
\left[ \varphi^{m}, H \right] & = & 2\varphi_{1}^{m} , \\
\left[ \varphi_{1}^{0}, H \right] & = & 
- \partial_{m}\varphi_{1}^{m} + \frac{m^{2}}{2}\varphi_{2}^{0} , 
\\
\left[ \varphi_{1}^{m}, H \right] & = & 
\frac{1}{2}\partial_{n}\left(\partial^{m}\varphi^{n} + 
\partial^{n}\varphi^{m}\right) 
+ \partial^{m}\varphi_{1}^{0} , \\
\left[ \varphi_{2}^{0}, H \right] & = & 
\partial_{m}\varphi^{m} + 4\varphi_{1}^{0} .
\end{eqnarray}
It is seen that $\varphi^{0}$ is a first-class constraint and the 
other four constraints are of the second class.

\section{Batalin-Fradkin extension}

In this section we convert the second-class constraints into 
first-class ones by applying the BF algorithm. It is seen that 
for both cases of $a \not= 1$  and $a=1$, the introduction of an 
auxiliary vector field (BF field) $\theta_{\mu}$ and its conjugate 
momentum $\omega^{\mu}$ is sufficient to modify the constraints 
and the Hamiltonian.

\subsection{The case of $a \not= 1$}
By the use of $\theta_{\mu}$ and $\omega^{\mu}$, we define new 
field variables as follows,
\begin{eqnarray}
\tilde{h}_{0} & \definition & 
h_{0} + \theta_{0} , \\
\tilde{h}_{m} & \definition & 
h_{m} + \theta_{m} 
- \frac{a}{1-a}\frac{1}{m^{2}}\partial_{m}\omega^{0} , \\
\tilde{h}_{mn} & \definition & 
h_{mn} - \frac{1}{2m^{2}}\left( 
\partial_{m}\omega_{n} + \partial_{n}\omega_{m}\right) , \\
\tilde{\pi}^{0} & \definition & 
\pi^{0} - \omega^{0} , \\
\tilde{\pi}^{m} & \definition & 
\pi^{m} - \omega^{m} - 
\frac{1}{2m^{2}}\partial_{n}\left( 
\partial^{m}\omega^{n} - \partial^{n}\omega^{m}\right) , \\
\tilde{\pi}^{mn} & \definition & 
\pi^{mn} + \frac{1}{2}\left(
\partial^{m}\theta^{n} + \partial^{n}\theta^{m}\right)
- \eta^{mn}\partial^{l}\theta_{l} 
- \frac{a}{1-a}\eta^{mn}\omega^{0} \nonumber \\
& & \makebox[2cm]{}
- \frac{1}{m^{2}}\left( \partial^{m}\partial^{n} 
- \eta^{mn}\partial_{l}\partial^{l}\right) \omega^{0} .
\end{eqnarray}
The constraints 
$\varphi^{A} \definition 
\left( \varphi^{0}, \varphi^{m}, \varphi_{1}^{0}, 
\varphi_{1}^{m}\right)$ and the Hamiltonian are modified as
\begin{eqnarray}
\tilde{\varphi}^{A} & \definition & 
\varphi^{A} \left[\left( h, \pi\right) \rightarrow 
\left(\tilde{h}, \tilde{\pi}\right)\right] 
\approx 0 , \label{eq:307}\\
\tilde{H} & \definition & 
H \left[\left( h, \pi\right) \rightarrow 
\left(\tilde{h}, \tilde{\pi}\right)\right] .\label{eq:308}
\end{eqnarray}
Their concrete forms are
\begin{eqnarray}
\tilde{\varphi}^{0} & = & 
\varphi^{0} - \omega^{0} \approx 0, \\
\tilde{\varphi}^{m} & = & 
\varphi^{m} - \omega^{m} \approx 0, \\
\tilde{\varphi}_{1}^{0} & = & 
\varphi_{1}^{0} - (1-a)m^{2}\theta_{0} + a \partial_{m}\omega^{m} 
\approx 0, \\
\tilde{\varphi}_{1}^{m} & = & 
\varphi_{1}^{m} + m^{2}\theta^{m} \approx 0,
\end{eqnarray}
and
\begin{eqnarray}
\tilde{H} & = &
H 
+ \left[\frac{1-2a}{1-a}\partial^{m}\theta_{m} 
+ \frac{2a}{1-a}\left(\frac{1}{m^{2}}\partial_{m}\partial^{m} 
- \frac{3a}{4(1-a)}\right)\omega^{0}\right]\tilde{\varphi}^{0} 
\nonumber \\ 
& & \makebox[5mm]{}
+ \left[ a\partial_{m}\theta_{0} 
- \frac{1}{2m^{2}}\partial^{n}\left(
(1-2a)\partial_{m}\omega_{n} + \partial_{n}\omega_{m} 
\right)\right]\tilde{\varphi}^{m} 
\nonumber \\ 
& & \makebox[5mm]{}
- \theta_{0}\tilde{\varphi}_{1}^{0} 
+ \left( -2\theta_{m} 
+ \frac{1-2a}{1-a}\frac{1}{m^{2}}\partial_{m}\omega^{0}\right)
\tilde{\varphi}_{1}^{m} \nonumber \\ 
& & \makebox[5mm]{}
- \frac{1-a}{2}m^{2}\theta_{0}^{\makebox[1mm]{}2} 
+ \frac{1-2a}{1-a}\omega^{0}\partial^{m}\theta_{m} 
+ m^{2}\theta_{m}\theta^{m} 
\nonumber \\ 
& & \makebox[5mm]{}
+ \frac{1}{4m^{2}}\left[
\partial_{m}\omega_{n}\partial^{m}\omega^{n} 
+ (1-2a)\left(\partial_{m}\omega^{m}\right)^{2}\right] 
\nonumber \\ 
& & \makebox[5mm]{}
- \frac{a}{1-a}\frac{1}{m^{2}}\partial_{m}\omega^{0}
\partial^{m}\omega^{0} 
- \frac{3a^{2}}{4(1-a)^{2}}\omega^{02} .
\end{eqnarray}
These modified set of constraints and Hamiltonian gives indeed a 
first-class constrained system:
\begin{eqnarray}
\left[ \tilde{\varphi}^{A}(\mib{x}), 
\tilde{\varphi}^{B}(\mib{x}')\right] = 0 ,\\
\left[ \tilde{\varphi}^{A}, \tilde{H}\right] = 0 .
\end{eqnarray}

\subsection{The case of $a=1$}

The results for this case have been given in I. 
We define the following new field variables:
\begin{eqnarray}
\tilde{h}_{0} & \definition & h_{0} , \\
\tilde{h}_{m} & \definition & 
h_{m} + \theta_{m} - \frac{1}{m^{2}}\partial_{m}\omega^{0} , \\
\tilde{h}_{mn} & \definition & 
h_{mn} - \frac{1}{2m^{2}}\left( 
\partial_{m}\omega_{n} + \partial_{n}\omega_{m}\right) - 
\frac{1}{3}\left(\eta_{mn} 
+ \frac{2}{m^{2}}\partial_{m}\partial_{n}\right) \theta_{0} , \\
\tilde{\pi}^{0} & \definition & \pi^{0} , \\
\tilde{\pi}^{m} & \definition & 
\pi^{m} - \omega^{m} - \frac{1}{2m^{2}}\partial_{n}\left( 
\partial^{m}\omega^{n} - \partial^{n}\omega^{m}\right) 
- \frac{2}{3}\partial^{m}\theta_{0} , \\
\tilde{\pi}^{mn} & \definition & 
\pi^{mn} + \frac{1}{2}\left(\partial^{m}\theta^{n} 
+ \partial^{n}\theta^{m}\right) 
- \eta^{mn}\partial^{l}\theta_{l} + \eta^{mn}\omega^{0} .
\end{eqnarray}
The modification of the constraints and the Hamiltonian is carried 
out as (\ref{eq:307}) and (\ref{eq:308}), which gives 
\begin{eqnarray}
\tilde{\varphi}^{0} & = & \varphi^{0} \approx 0 , \\
\tilde{\varphi}^{m} & = & \varphi^{m} - \omega^{m} 
\approx 0 , \\
\tilde{\varphi}_{1}^{0} & = & 
\varphi_{1}^{0} + \partial_{m}\omega^{m} + m^{2}\theta_{0} 
\approx 0 , \\
\tilde{\varphi}_{1}^{m} & = & 
\varphi_{1}^{m} + m^{2}\theta^{m} 
\approx 0 , \\
\tilde{\varphi}_{2}^{0} & = & 
\varphi_{2}^{0} - 2\partial^{m}\theta_{m} + 3\omega^{0} 
\approx 0 ,
\end{eqnarray}
and 
\begin{eqnarray}
\tilde{H} & = & H + 
\left[ 
- \frac{1}{3}\left( 1 + \frac{2}{m^{2}}\partial_{n}\partial^{n}
\right)\partial_{m}\theta_{0} - 
\frac{1}{2m^{2}}\partial^{n}\left( 
\partial_{m}\omega_{n} + \partial_{n}\omega_{m}\right)\right]
\tilde{\varphi}^{m} \nonumber \\
& & \makebox[5mm]{}
+ \left[\frac{1}{3}\left( 4 
+ \frac{2}{m^{2}}\partial_{m}\partial^{m}\right)\theta_{0} 
+ \frac{1}{m^{2}}\partial_{m}\omega^{m}\right]
\tilde{\varphi}_{1}^{0} \nonumber \\
& & \makebox[5mm]{}
+ \left( 
- 2\theta_{m} + \frac{1}{m^{2}}\partial_{m}\omega^{0}\right)
\tilde{\varphi}_{1}^{m}
- \frac{1}{2}\omega^{0}\tilde{\varphi}_{2}^{0} 
\nonumber \\
& & \makebox[5mm]{} + 
\frac{1}{3}\partial_{m}\theta_{0}\partial^{m}\theta_{0} - 
\frac{2}{3}m^{2}\theta_{0}^{\makebox[1mm]{}2} - 
\theta_{0}\partial_{m}\omega^{m} + 
\frac{1}{8m^{2}}\left( 
\partial_{m}\omega_{n} - \partial_{n}\omega_{m}\right)^{2} 
\nonumber \\
& & \makebox[5mm]{} + m^{2}\theta_{m}\theta^{m} + 
\frac{3}{4}\omega^{02} .
\end{eqnarray}

\section{Gauge fixing}

\subsection{The case of $a \not= 1$}
In order to find a massless-regular theory, we impose the following 
gauge-fixing conditions 
$\chi_{A} \definition 
\left(\chi_{0}, \chi_{m}, \chi_{10}, \chi_{1m} \right)$:
\begin{eqnarray}
\chi_{0} & \definition & h_{0} \approx 0 , \\
\chi_{m} & \definition & h_{m} \approx 0 , \\
\chi_{10} & \definition & 
\pi' - \frac{3a}{1-a}\omega^{0} \approx 0 , \\
\chi_{1m} & \definition & 
\partial^{n}h_{mn} - \frac{1}{2}\partial_{m}h' \approx 0 .
\end{eqnarray}
The path integral is given by
\begin{eqnarray}
Z & = & 
\int{\cal D}\pi^{0}{\cal D}\pi^{m}{\cal D}\pi^{mn}{\cal D}h_{0}
{\cal D}h_{m}{\cal D}h_{mn}{\cal D}\omega^{0}{\cal D}\omega^{m}
{\cal D}\theta_{0}{\cal D}\theta_{m}
\prod_{A}
\delta(\tilde{\varphi}^{A})\delta(\chi_{A})\prod_{t}{\rm Det}M 
\nonumber \\
& & \makebox[1cm]{} \times
\exp i\int d^{4}x\left[
\pi^{0}\dot{h}_{0} + \pi^{m}\dot{h}_{m} + \pi^{mn}\dot{h}_{mn} + 
\omega^{0}\dot{\theta}_{0} + \omega^{m}\dot{\theta}_{m} - \tilde{H}
\right] ,
\end{eqnarray}
where
\begin{equation}
M \definition \delta_{\alpha}^{\beta}\partial_{m}\partial^{m}
\delta^{3}( \mib{x} - \mib{x}' ) .
\makebox[1cm]{}( \alpha, \beta = 0-3 )
\end{equation}
The integrations over $\pi^{0}, \pi^{m}, h_{0}$ and $h_{m}$ 
are easily carried out. The $\delta$-functions 
$\delta(\tilde{\varphi}_{1}^{0}), 
\delta(\tilde{\varphi}_{1}^{m})$ and 
$\delta(\chi_{10})$ are exponentiated as
\begin{equation}
\delta(\tilde{\varphi}_{1}^{0})
\delta(\tilde{\varphi}_{1}^{m})
\delta(\chi_{10}) = 
\int{\cal D}\lambda_{0}{\cal D}\lambda_{m}{\cal D}\mu
\exp i\int d^{4}x\left[
\lambda_{0}\tilde{\varphi}_{1}^{0} + 
\lambda_{m}\tilde{\varphi}_{1}^{m} + 
\mu\chi_{10}\right] .
\end{equation}
We integrate over $\pi^{mn}$, write $2h_{m}$ and $h_{0}$ 
over $\lambda_{m}$ and $\lambda_{0}$ respectively, and 
further integrate with respect to $\theta_{0}, \theta_{m}$ 
and $\mu$. 
Replacing variables as 
$\omega^{m} \rightarrow 2m^{2}\theta_{m}$ and 
$\omega^{0} \rightarrow - 2(1-a)m^{2}\theta_{0}$, 
we obtain
\begin{eqnarray}
Z & = & 
\int{\cal D}h_{\mu\nu}{\cal D}\theta_{\mu}
\delta\left(
\partial^{n}h_{mn} - \frac{1}{2}\partial_{m}h'\right)
\prod_{t}{\rm Det}M
\nonumber \\
& & \makebox[2cm]{}
\times\exp i \int d^{4}x
\left[ L[ h, \theta ] 
+ \frac{4}{3}\left(\partial^{m}h_{m} 
- \frac{1}{2}\dot{h}'\right)^{2}\right] , \label{eq:408}
\end{eqnarray}
where
\begin{equation}
L[ h, \theta ] \definition 
L[ h, m=0 ] - 
\frac{m^{2}}{2}\left(
\left(h_{\mu\nu} - \partial_{\mu}\theta_{\nu} 
- \partial_{\nu}\theta_{\mu}\right)^{2} 
- a\left( h - 2\partial^{\mu}\theta_{\mu}\right)^{2}\right) . 
\label{eq:409}
\end{equation}

\subsection{The case of $a=1$}

For this case, We imposed the following gauge-fixing conditions in I.
\begin{eqnarray}
\chi_{0} & \definition & h_{0} \approx 0 , \\
\chi_{m} & \definition & h_{m} \approx 0 , \\
\chi_{10} & \definition & \pi' + 3\omega^{0} \approx 0 , \\
\chi_{1m} & \definition & 
\partial^{n}h_{mn} - \frac{1}{2}\partial_{m}h' \approx 0 , \\
\chi_{20} & \definition & \theta_{0} \approx 0 .
\end{eqnarray}
The final expression obtained in I was 
\begin{eqnarray}
Z & = & 
\int{\cal D}h_{\mu\nu}{\cal D}\theta_{\mu}
\delta\left(
\partial^{m}h_{m} - \frac{1}{2}\dot{h}'\right)
\delta\left(
\partial^{n}h_{mn} - \frac{1}{2}\partial_{m}h'\right)
\prod_{t}{\rm Det}M \nonumber \\
& & \makebox[2cm]{}
\times\exp i \int d^{4}x
\left[ L[ h, \theta ] 
+ 4m^{2}\dot{\theta}_{0}\left(
\partial^{m}\theta_{m} - \frac{1}{2}h'\right)\right] , 
\label{eq:415}
\end{eqnarray}
where
\begin{equation}
L[ h, \theta ] \definition 
L[ h, m=0 ] - 
\frac{m^{2}}{2}\left(\left(
h_{\mu\nu} - \partial_{\mu}\theta_{\nu} 
- \partial_{\nu}\theta_{\mu}\right)^{2} - 
\left(h - 2\partial^{\mu}\theta_{\mu}\right)^{2}\right) .
\end{equation}

\section{Covariant expressions}

\subsection{The case of $a \not= 1$}

It is easy to obtain covariant expressions for the generating 
functional $Z$ (\ref{eq:408}). 
In consideration of the fact that the Lagrangian $L[ h, \theta ]$ 
is invariant under the gauge transformation with four arbitrary 
functions $\varepsilon_{\mu}(x)$ 
\begin{equation}
\left\{
\begin{array}{lll}
\delta_{\varepsilon}h_{\mu\nu} & = & 
\partial_{\mu}\varepsilon_{\nu} + \partial_{\nu}\varepsilon_{\mu} ,
\\
\delta_{\varepsilon}\theta_{\mu} & = & \varepsilon_{\mu} ,
\end{array}
\right.
\end{equation}
we can give various expressions for $Z$. 
The situation is almost the same as in a massless tensor field. 
For example, for a `Coulomb-like gauge', we have 
\begin{eqnarray}
Z & = & 
\int{\cal D}h_{\mu\nu}{\cal D}\theta_{\mu}
\delta\left(
\partial^{m}h_{m} - \frac{1}{2}\dot{h}' - f_{0}\right)
\delta\left(
\partial^{n}h_{mn} - \frac{1}{2}\partial_{m}h' - f_{m}\right)
\prod_{t}{\rm Det}M \nonumber \\
& & \makebox[2cm]{} \times\exp i \int d^{4}x L[ h, \theta ] \\
& = & 
\int{\cal D}h_{\mu\nu}{\cal D}\theta_{\mu}
\delta\left(
\partial^{n}h_{mn} - \frac{1}{2}\partial_{m}h' - f_{m}\right)
\prod_{t}{\rm Det}M \nonumber \\
& & \makebox[2cm]{} \times\exp i \int d^{4}x
\left[ L[ h, \theta ] 
+ \frac{1}{2\alpha}\left(\partial^{m}h_{m} - \frac{1}{2}\dot{h}'\right)^{2}\right] , \label{eq:503}
\end{eqnarray}
where $f_{\mu}(\mu = 0 - 3)$ are arbitrary functions of $x$, 
and $\alpha$ is an arbitrary constant, gauge parameter. 
The expression (\ref{eq:408}) is a special case of (\ref{eq:503}).
The covariant expressions are also obtained as follows:
\begin{eqnarray}
Z & = &
\int{\cal D}h_{\mu\nu}{\cal D}\theta_{\mu}
\delta(\partial^{\nu}h_{\mu\nu} - 
\frac{1}{2}\partial_{\mu}h - f_{\mu})
{\rm Det}N\exp i\int d^{4}x L[h, \theta ] \\
& = &
\int{\cal D}h_{\mu\nu}{\cal D}\theta_{\mu}{\cal D}B_{\mu}
{\cal D}c_{\mu}{\cal D}\bar{c}^{\mu}
\nonumber \\
& & \makebox[1cm]{}
\times\exp i\int d^{4}x\left[ L[h, \theta ] + 
B^{\mu}\left(\partial^{\nu}h_{\mu\nu} - 
\frac{1}{2}\partial_{\mu}h + 
\frac{\alpha}{2}B_{\mu}\right) + 
i\bar{c}^{\mu}\Box c_{\mu}\right] , \nonumber \\
& &
\end{eqnarray}
where $N$ is defined by
\begin{equation}
N \definition \delta_{\alpha}^{\beta}\Box\delta^{4}(x-x') , 
\makebox[2cm]{}( \alpha , \beta = 0-3 )
\end{equation}
and the Nakanishi-Lautrup (NL) field $B_{\mu}$ and the 
Faddeev-Popov (FP) ghosts $( c_{\mu}, \bar{c}^{\mu} )$ have been 
introduced.

\subsection{The case of $a=1$}

As the first step to obtain covariant expressions, 
we introduce another auxiliary field of scalar type $\varphi(x)$ 
and define a gauge transformation with five arbitrary functions 
$\varepsilon_{\mu}(x)$ and $\varepsilon(x)$:
\begin{equation}
\left\{
\begin{array}{lll}
\delta_{\varepsilon}h_{\mu\nu} & = & 
\partial_{\mu}\varepsilon_{\nu} + \partial_{\nu}\varepsilon_{\mu} ,
\\
\delta_{\varepsilon}\theta_{\mu} & = & 
\varepsilon_{\mu} + \partial_{\mu}\varepsilon ,\\
\delta_{\varepsilon}\varphi & = & \varepsilon .
\end{array}
\right.
\end{equation}
The expression (\ref{eq:415}) can be written as
\begin{eqnarray}
Z & = & 
\int{\cal D}h_{\mu\nu}{\cal D}\theta_{\mu}{\cal D}\varphi
\delta\left(
\partial^{m}h_{m} - \frac{1}{2}\dot{h}'\right)
\delta\left(
\partial^{n}h_{mn} - \frac{1}{2}\partial_{m}h'\right)
\delta\left(\partial_{m}\partial^{m}\varphi\right)
\prod_{t}{\rm Det}M'
\nonumber \\
& & \makebox[2cm]{} \times\exp i \int d^{4}x
\left[ L[ h, \theta ] 
+ 4m^{2}\dot{\theta}_{0}\left(
\partial^{m}\theta_{m} - \frac{1}{2}h'\right)\right] , 
\label{eq:508}
\end{eqnarray}
where
\begin{equation}
M' \definition \delta_{\alpha}^{\beta}\partial_{m}\partial^{m}
\delta^{3}( \mib{x} - \mib{x}' ) .
\makebox[1cm]{}( \alpha, \beta = 0-4 )
\end{equation}
The Lagrangian $L[ h, \theta ]$ can be decomposed into a manifestly 
gauge-invariant part $L[ h, \theta, \varphi ]$ and the 
rest $R[ h, \theta, \varphi ]$ as
\begin{equation}
L[ h, \theta ] = 
L[ h, \theta, \varphi ] + R[ h, \theta, \varphi ] ,
\end{equation}
where
\begin{eqnarray}
L[ h, \theta, \varphi ] & \definition & 
L[h, m=0] - \frac{m^{2}}{2}\left[\left(
h_{\mu\nu} - \partial_{\mu}\theta_{\nu} 
- \partial_{\nu}\theta_{\mu}
+ 2\partial_{\mu}\partial_{\nu}\varphi\right)^{2}\right. 
\nonumber \\ 
& & \makebox[35mm]{} \left. - \left(
h - 2\partial^{\mu}\theta_{\mu}+ 2\Box\varphi\right)^{2}\right] , 
\\
R[ h, \theta, \varphi ] & \definition & 
4m^{2}\left[
\left(\partial^{m}h_{m} - \frac{1}{2}\dot{h}'\right)\dot{\varphi} 
- \frac{1}{2}\left(\partial^{n}h_{mn} 
- \frac{1}{2}\partial_{m}h'\right)\partial^{m}\varphi \right.
\nonumber \\
& & \makebox[2cm]{}
+ \left.\frac{1}{2}\left( h_{0} - \frac{1}{2}h'\right)
\partial_{m}\partial^{m}\varphi\right] .
\end{eqnarray}
Because of the existence of three $\delta$-functions 
in (\ref{eq:508}), $R$ has null effect.
The path integral then reduces to
\begin{eqnarray}
Z & = & 
\int{\cal D}h_{\mu\nu}{\cal D}\theta_{\mu}{\cal D}\varphi
\delta\left(
\partial^{m}h_{m} - \frac{1}{2}\dot{h}'\right)
\delta\left(
\partial^{n}h_{mn} - \frac{1}{2}\partial_{m}h'\right)
\delta\left(\partial_{m}\partial^{m}\varphi\right)
\prod_{t}{\rm Det}M'
\nonumber \\
& & \makebox[2cm]{} \times\exp i \int d^{4}x\left[
L[ h, \theta, \varphi ] 
+ 4m^{2}\dot{\theta}_{0}\left(\partial^{m}\theta_{m} 
- \frac{1}{2}h'\right)\right] . \label{eq:513}
\end{eqnarray}

As the next step, we define the following two gauge-invariant 
quantities:
\begin{eqnarray}
\Delta^{-1}[ h, \theta, \varphi ] & \definition & 
\int{\cal D}\varepsilon
\delta\left(\partial^{m}h^{\varepsilon}_{m} - 
\frac{1}{2}\dot{h}'^{\varepsilon}\right)
\delta\left(\partial^{n}h^{\varepsilon}_{mn} - 
\frac{1}{2}\partial_{m}h'^{\varepsilon}\right)
\delta\left(\partial_{m}\partial^{m}\varphi^{\varepsilon}\right) ,
\nonumber \\
& & \makebox[2cm]{} \times
\exp i\int d^{4}x\left[
4m^{2}\dot{\theta}_{0}^{\varepsilon}\left(
\partial^{m}\theta^{\varepsilon}_{m} - 
\frac{1}{2}h'^{\varepsilon}\right)\right] , \\
\Delta '^{-1}[ h, \theta, \varphi ] & \definition &
\int{\cal D}\varepsilon
\delta\left(\partial^{m}h^{\varepsilon}_{m} - 
\frac{1}{2}\dot{h}'^{\varepsilon}\right)
\delta\left(\partial^{n}h^{\varepsilon}_{mn} - 
\frac{1}{2}\partial_{m}h'^{\varepsilon}\right)
\delta\left( \partial^{m}\theta^{\varepsilon}_{m}
- \frac{1}{2}h'^{\varepsilon}\right) . \nonumber \\
& & 
\label{eq:515}
\end{eqnarray}
In the above, ${\cal D}\varepsilon$ stands for the invariant 
measure on the gauge group, having the property
\begin{equation}
{\cal D}\varepsilon = {\cal D}(\varepsilon\varepsilon') 
= {\cal D}(\varepsilon'\varepsilon) = {\cal D}\bar{\varepsilon} ,
\end{equation}
where $\bar{\varepsilon}$ denotes the inverse of $\varepsilon$.
The quantities like $h^{\varepsilon}$ indicate 
gauge-transformed ones of the respective fields.
For $\Delta'[ h, \theta, \varphi ]$ we simply have
\begin{equation}
\Delta'[ h, \theta, \varphi ] = \prod_{t}{\rm Det}M' .
\end{equation}
To evaluate $\Delta[ h, \theta, \varphi ]$, we take a 
special gauge orbit ${\cal O}$ that contains a configuration 
$(h,\theta,\varphi)$ satisfying
\begin{equation}
\left\{
\begin{array}{rcl}
\displaystyle 
\partial^{m}h_{m} - \frac{1}{2}\dot{h}' & = & 0 ,
\vspace{2mm} \\
\displaystyle 
\partial^{n}h_{mn} - \frac{1}{2}\partial_{m}h' & = & 0 ,
\vspace{2mm} \\
\displaystyle
\partial^{m}\theta_{m} - \frac{1}{2}h' & = & 0 .
\end{array}
\right.
\end{equation}
For this configuration the quantity $\Delta[ h, \theta, \varphi ]$ 
is calculated as
\begin{eqnarray}
\Delta^{-1}[ h, \theta, \varphi ] & = & 
\int{\cal D}\varepsilon
\delta\left(\partial_{m}\partial^{m}\varepsilon_{0}\right)
\delta\left(\partial_{n}\partial^{n}\varepsilon_{m}\right)
\delta\left(\partial_{m}\partial^{m}\varphi
+ \partial_{m}\partial^{m}\varepsilon\right)
\nonumber \\
& & \makebox[2cm]{} \times
\exp i\int d^{4}x\left[
4m^{2}\left(\dot{\theta}_{0} + \dot{\varepsilon}_{0} 
+ \ddot{\varepsilon}\right)\partial_{m}\partial^{m}\varepsilon
\right] \nonumber \\
& = &
\int{\cal D}\varepsilon
\delta\left(\partial_{m}\partial^{m}\varepsilon_{0}\right)
\delta\left(\partial_{n}\partial^{n}\varepsilon_{m}\right)
\delta\left(\partial_{m}\partial^{m}\varphi
+ \partial_{m}\partial^{m}\varepsilon\right)
\nonumber \\
& & \makebox[2cm]{} \times
\exp i\int d^{4}x\left[
- 4m^{2}\left(\dot{\theta}_{0} - \ddot{\varphi}\right)
\partial_{m}\partial^{m}\varphi\right] 
\nonumber \\
& = &
\left(\prod_{t}{\rm Det}M'\right)^{-1}
\exp i\int d^{4}x\left[
- 4m^{2}\left(\dot{\theta}_{0} - \ddot{\varphi}\right)
\partial_{m}\partial^{m}\varphi\right] . \label{eq:519}
\end{eqnarray}
The gauge invariance of this quantity tells that the expression 
(\ref{eq:519}) is valid for any configuration belonging to 
${\cal O}$. 
Since the configuration $(\bar{h},\bar{\theta},\bar{\varphi})$ 
such that
\begin{equation}
\left\{
\begin{array}{rcl}
\displaystyle 
\partial^{m}\bar{h}_{m} - \frac{1}{2}\dot{\bar{h}}' & = & 0 ,
\vspace{2mm} \\
\displaystyle 
\partial^{n}\bar{h}_{mn} - \frac{1}{2}\partial_{m}\bar{h}' 
& = & 0 ,
\vspace{2mm} \\
\displaystyle
\partial_{m}\partial^{m}\bar{\varphi} & = & 0 
\end{array}
\right.
\end{equation}
belongs to ${\cal O}$, we can evaluate $\Delta[h,\theta,\varphi]$ 
for this configuration. We then have
\begin{equation}
\Delta[h,\theta,\varphi] = 
\Delta[\bar{h},\bar{\theta},\bar{\varphi}] = \prod_{t}{\rm Det}M' .
\label{eq:521}
\end{equation}
Multiply Eq. (\ref{eq:513}) by 
\[
[\mbox{The right hand side of (\ref{eq:515})}]
\times\Delta'[ h, \theta, \varphi ]
=1
\]
to give
\begin{eqnarray}
Z & = & 
\int{\cal D}h_{\mu\nu}{\cal D}\theta_{\mu}{\cal D}\varphi
\delta\left(
\partial^{m}h_{m} - \frac{1}{2}\dot{h}'\right)
\delta\left(
\partial^{n}h_{mn} - \frac{1}{2}\partial_{m}h'\right)
\delta\left(\partial_{m}\partial^{m}\varphi\right)
\prod_{t}{\rm Det}M'
\nonumber \\
& & \times\exp i \int d^{4}x\left[
L[ h, \theta, \varphi ] 
+ 4m^{2}\dot{\theta}_{0}\left(\partial^{m}\theta_{m} 
- \frac{1}{2}h'\right)\right] 
\nonumber \\
& & \times
\int{\cal D}\varepsilon
\delta\left(\partial^{m}h^{\varepsilon}_{m} - 
\frac{1}{2}\dot{h}'^{\varepsilon}\right)
\delta\left(\partial^{n}h^{\varepsilon}_{mn} - 
\frac{1}{2}\partial_{m}h'^{\varepsilon}\right)
\delta\left( \partial^{m}\theta^{\varepsilon}_{m}
- \frac{1}{2}h'^{\varepsilon}\right) 
\Delta '[ h, \theta, \varphi ] .
\nonumber \\
& &
\end{eqnarray}
Change variables from $( h, \theta, \varphi )$ to 
$( h', \theta', \varphi' ) \definition ( h^{\varepsilon}, 
\theta^{\varepsilon}, \varphi^{\varepsilon} )$, 
take into account the gauge-invariance of 
$L[ h, \theta, \varphi ]$ and $\Delta'[ h, \theta, \varphi ]$ 
as well as that of the measure 
${\cal D}h_{\mu\nu}{\cal D}\theta_{\mu}{\cal D}\varphi$, and remove 
prime signs.
Then we have
\begin{eqnarray}
Z & = & 
\int{\cal D}h_{\mu\nu}{\cal D}\theta_{\mu}{\cal D}\varphi 
\nonumber \\
& & \times
\left\{\left[
\int{\cal D}\bar{\varepsilon}
\delta\left(
\partial^{m}h^{\bar{\varepsilon}}_{m} 
- \frac{1}{2}\dot{h}'^{\bar{\varepsilon}}\right)
\delta\left(
\partial^{n}h^{\bar{\varepsilon}}_{mn} 
- \frac{1}{2}\partial_{m}h'^{\bar{\varepsilon}}\right)
\delta\left(
\partial_{m}\partial^{m}\varphi^{\bar{\varepsilon}}\right)
\right.\right. \nonumber \\
& & \makebox[2cm]{}
\left.\left.
\times\exp i\int d^{4}x\left(
4m^{2}\dot{\theta}^{\bar{\varepsilon}}_{0}
\left(\partial^{m}\theta^{\bar{\varepsilon}}_{m} 
- \frac{1}{2}h'^{\bar{\varepsilon}}\right)\right)\right]
\prod_{t}{\rm Det}M'\right\}
\nonumber \\
& & \times
\exp i \int d^{4}xL[ h, \theta, \varphi ]
\nonumber \\
& & \times
\delta\left(\partial^{m}h_{m} - \frac{1}{2}\dot{h}'\right)
\delta\left(\partial^{n}h_{mn} - \frac{1}{2}\partial_{m}h'\right)
\delta\left(\partial^{m}\theta_{m} - \frac{1}{2}h'\right)
\Delta'[h, \theta, \varphi ] .
\nonumber \\
& & 
\end{eqnarray}
In this expression, the factor enclosed in braces is equal to 1 
as seen from (\ref{eq:521}).
The path integral can be expressed as follws, 
\begin{eqnarray}
Z & = & 
\int{\cal D}h_{\mu\nu}{\cal D}\theta_{\mu}{\cal D}\varphi
\delta\left(\partial^{m}h_{m} - \frac{1}{2}\dot{h}'\right)
\delta\left(\partial^{n}h_{mn} - \frac{1}{2}\partial_{m}h'\right)
\delta\left(\partial^{m}\theta_{m} - \frac{1}{2}h'\right)
\nonumber \\
& & \makebox[2cm]{} \times\prod_{t}{\rm Det}M'
\exp i\int d^{4}x L[ h, \theta, \varphi ] .
\end{eqnarray}

Coming to this stage, it is an easy task to give covariant forms to 
$Z$.
That is
\begin{eqnarray}
Z & = & 
\int{\cal D}h_{\mu\nu}{\cal D}\theta_{\mu}{\cal D}\varphi
\delta\left(\partial^{\nu}h_{\mu\nu} - 
\frac{1}{2}\partial_{\mu}h - f_{\mu}\right)
\delta\left(\partial^{\mu}\theta_{\mu} - \frac{1}{2}h - f\right)
{\rm Det}N'
\nonumber \\
& & \makebox[2cm]{} \times
\exp i\int d^{4}x L[ h, \theta, \varphi ] ,
\end{eqnarray}
where $N'$ is defined by
\begin{equation}
N' \definition \delta_{\alpha}^{\beta}\Box\delta^{4}(x-x') , 
\makebox[2cm]{}( \alpha , \beta = 0-4 )
\end{equation}
and $f_{\mu}$ and $f$ are arbitrary functions of $x$.
By introducing the Nakanishi-Lautrup fields $( B_{\mu}, B )$ and the 
Faddeev-Popov ghosts $( c_{\mu}, c )$ and 
$( \bar{c}^{\mu}, \bar{c} )$, 
we arrive at the final form of the path integral
\begin{eqnarray}
Z & = & 
\int{\cal D}h_{\mu\nu}{\cal D}\theta_{\mu}{\cal D}\varphi
{\cal D}B_{\mu}{\cal D}B{\cal D}c_{\mu}{\cal D}\bar{c}^{\mu}
{\cal D}c{\cal D}\bar{c}
\nonumber \\
& & \makebox[2cm]{} \times
\exp i\int d^{4}x\left[ L[ h, \theta, \varphi ]
+ L_{{\rm GF+FP}}
\right] ,
\end{eqnarray}
\begin{eqnarray}
L_{{\rm GF+FP}} & = &
B^{\mu}\left(\partial^{\nu}h_{\mu\nu} 
- \frac{1}{2}\partial_{\mu}h + 
\frac{\alpha}{2}B_{\mu}\right) + 
i\bar{c}^{\mu}\Box c_{\mu}
\nonumber \\
& & \makebox[2cm]{}
+ B\left(\partial^{\mu}\theta_{\mu} - \frac{1}{2}h 
+ \frac{\beta}{2}B\right) +
i\bar{c}\Box c ,
\end{eqnarray}
where $\alpha$ and $\beta$ are arbitrary constants, gauge parameters.

\section{Summary}

We have given the covariant path integral expressions to the gauge 
theories of massive tensor fields. It has turned out that in the 
case of the PT model a scalar field in addition to a vector field 
has to be introduced as auxiliary BF field, while only a vector 
field is  necessary for the ASG model. The difference comes from 
that of the constraint structures in the two models.

To construct a complete nonlinear theory which smoothly reduces to 
general relativity in the massless limit is left for future study.

\section*{Acknowledgements}
We would like to thank T.~Kurimoto for discussion.



\end{document}